\pdfoutput=1
\documentclass[review]{elsarticle}

\usepackage{lineno,hyperref}
\modulolinenumbers[5]

\journal{Discrete Applied Mathematics}

\RequirePackage{subcaption}
\RequirePackage{amsmath}

\def\cP{P}
\def\cH{\mathcal{H}}
\def\cS{S}

\def\argmin{\mathop{arg\,min}\limits}

\RequirePackage[ruled]{algorithm}
\RequirePackage[noend]{algorithmic}

\newcommand{\algKeyword}[1]{{\bf #1}}
\newcommand{\Proc}[1]{\text{\tt #1}}
\def\CALL{\algKeyword{call}~}

\newenvironment{AlgProcedure}[1]
{
    \small
    \medskip
    \medskip
    \algKeyword{PROCEDURE} #1
    \begin{algorithmic}[1]}
    {\end{algorithmic}
    \bigskip
}

\def\CALL{\algKeyword{call}~}

\begin{document}

\begin{frontmatter}

\title{Parallelizing asymptotically optimal algorithms for large-scale dualization
    problems}

\author{Djukova E.V.}
\ead{edjukova@mail.ru}

\author{Nikiforov A.G.}
\ead{ankifor@gmail.com}

\author{Prokofyev P.A.}
\ead{p\_prok@mail.ru}

\address{Vavilov Street, Moscow, 40}

\begin{abstract}
Dualization is a key discrete enumeration problem.
It is not known whether or not this problem is polynomial-time solvable.
Asymptotically optimal dualization algorithms are the fastest among the known dualization 
algorithms,
which is supported by new experiments with various data described in this paper.
A theoretical justification of the efficiency of these algorithms on the average was 
given by E.V. Djukova more than 30 years ago.
In this paper, new results on the construction of parallel algorithms for intractable 
enumeration problems are presented.
A new static parallelization scheme for asymptotically optimal dualization algorithms is 
developed and tested.
The scheme is based on statistical estimations of subtasks size.
\end{abstract}


\end{frontmatter}


\section{Introduction}

We consider dualization, which is the problem of searching for
irreducible coverings of a Boolean matrix. Let $L = \| a_{ij}
\|_{m \times n}$ be a Boolean matrix and $H$ be a set of columns
of~$L$. The set $H$ is called a \emph{covering} of~$L$ if each row
of~$L$ has at least one unit element in the columns~$H$. A
covering~$H$ is called \emph{irreducible} if any proper subset
of~$H$ is not a covering of~$L$. Let $\cP(L)$ denote the set of
all possible irreducible coverings of~$L$. The problem is to
construct $\cP(L)$.

There are other formulations of dualization, specifically, based
on concepts of the theory of Boolean functions and graph and
hypergraph theory. Let us present these formulations.

\begin{enumerate}
    \item[(1)]
    Given a conjunctive normal form consisting of~$m$ different clauses that implements a 
    monotone Boolean function $F(x_1,\ldots,x_n)$,
    construct a reduced disjunctive normal form of~$F$.
    \item[(2)]
    Given a hypergraph~$\cH$  consisting of~$n$  vertices and~$m$ edges, find all minimal 
    vertex coverings of~$\cH$.
\end{enumerate}

The efficiency of enumeration algorithms is characterized by the
complexity of a single step (see \cite{JYP1988}). An algorithm has
a (quasi-)polynomial-time delay if, for any individual problem,
each step of this algorithm (the construction of the current
solution) is executed in (quasi-)polynomial time in the input size
of the problem. As applied to the search for irreducible
coverings, this means that, for any $m{\times}n$ Boolean matrix,
the time required for the construction of the next irreducible
covering is bounded by a (quasi-)polynomial in~$m$ and~$n$. In the
general case no dualization algorithm with a (quasi-)polynomial
time delay has yet been constructed and it is not known whether
such an algorithm exists. There are examples of such algorithms
for some special cases of dualization \cite{JYP1988, EGM2003}. For
example, in \cite{JYP1988} an algorithm with a time delay $O( n^3
)$ was constructed in the case when each row of~$L$ has at most
two unit elements (in this case $\cH$ is a graph in formulation
(2)).

Studies concerning the complexity of enumeration problems
basically address the possibility of constructing incremental
(quasi-)polynomial-time algorithms. In this case, the incremental
property means that at every step in the construction of the
current solution an algorithm searches through the set of
solutions obtained at the preceding steps and the time taken by
this search is (quasi-)polynomial in the input problem size and
the number of previously found solutions. An incremental
quasi-polynomial-time dualization algorithm was constructed in
\cite{FK1996,KBEG2006}. For several special cases of dualization,
incremental polynomial-time algorithms were constructed in
\cite{BGEK2000,BE2004a}.

Another approach to the solution of the problem is based on the
concept of asymptotically optimal algorithm with a polynomial time
delay. This approach was first proposed in \cite{D1977} and deals
with a typical case.

According to this approach, the original enumeration problem  $Z$
is replaced by a УsimplerФ problem  $Z_1$ that has the same input
and is solved with a polynomial time delay. The solution set of
$Z_1$ contains the solution set of $Z$ and, second, with
increasing input size, the number of solutions of $Z_1$ is almost
always asymptotically equal to the number of solutions of $Z$.
This approach is substantiated by obtaining asymptotics for the
typical number of solutions to each of the problems~$Z$~and~$Z_1$.

Thus, in contrast to an "exact" algorithm with a polynomial time
delay, an asymptotically optimal algorithm can execute redundant
polynomial-time steps. A redundant step is a solution of $Z_1$
that was either found previously or is constructed for the first
time but is not a solution to the problem $Z$. For almost all
problems of a given size, the number of redundant steps must have
a lower order of growth than the number of all steps of the
algorithm as the problem input size increases. Whether or not a
step is redundant must be verifiable in a polynomial amount of
time in the problem input size.

A number of asymptotically optimal algorithms for constructing
irreducible coverings of a Boolean matrix have been proposed in
the case when the input matrix satisfies the condition $\log m \le
(1-\epsilon) \log n$, $0 < \epsilon < 1$, \cite{D1977, D2004,
    DI2008, D1987, DZ1997, DZ2000, D2003, DP2014, KA2010}. The
following criterion called USM is used to construct  $\cP( L )$ in
these algorithms. A~set  $H$ of  $r$ columns of the matrix $L$ is
an irreducible covering if and only if the following two
conditions hold:
\begin{enumerate}
    \item[(a)]
    the submatrix  $L^H$ of  $L$ made up of the columns
    of  $H$ does not contain rows of the form  $(0, 0, \ldots, 0)$ and
    \item[(b)] $L^H$ contains every row of the form $(1, 0, 0, \ldots,
    0, 0), (0, 1, 0, \ldots, 0, 0), \ldots ,\linebreak (0, 0, 0,
    \ldots, 0, 1)$;
    i.e., it contains the identity submatrix of order $r$.
\end{enumerate}
A set of columns satisfying condition (b) is called
\emph{consistent}. A consistent set of columns is called
\emph{maximal} if it is not contained in any other consistent set
of columns.

In the asymptotically optimal dualization algorithm AO1 (see
\cite{D1977}), $Z_1$ is the problem of constructing a collection
of column sets a matrix $L$ satisfying condition (b) in which each
set of length $r$ occurs as many times as there are identity
submatrices of order $r$ in this set. In fact, all identity
submatrices of L are enumerated with a polynomial time delay.
Clearly, an irreducible covering can be generated only by a
maximal identity submatrix, i.e., by an identity submatrix that is
not contained in any other one. A maximal identity submatrix
generates a maximal consistent set of columns, i.e., a consistent
set of columns that is not contained in any other one.

According to the algorithm AO1, the maximal identity submatrices
(the maximal consistent sets of columns) can be enumerated
(enumerated with repetition) with a step complexity of $O(qmn)$,
where $q = \min\{m, n\}$. As a result of enumerating the identity
submatrices, some sets of columns are repeatedly constructed. When
obtaining the current maximal identity submatrix $Q$ in time
$O(mn)$, the algorithm AO1 checks condition (a) for the set $H$ of
columns of $L$ generated by the submatrix $Q$. If condition (a)
holds, then AO1 checks in time $O(mn)$ whether $H$ was constructed
at a previous step.

The algorithm AO2 \cite{D2004}, which is a modification of AO1,
enumerates (with a polynomial time delay $O(qm^2n)$) only identity
submatrices of $L$ that generate coverings. At every step, AO2
constructs an irreducible covering. However, as in AO1, the
solutions can repeat. This algorithm takes less redundant steps
than AO1. Based on AO2, the algorithms AO2K and AO2M with a
reduced execution time were constructed in \cite{DP2014}.

The asymptotically optimal algorithm OPT enumerates without
repetitions and
with a polynomial time delay $O(qm^2n)$  the sets of columns of $L$
satisfying condition (b) and some additional conditions, including the maximality one
\cite{DI2008}. Redundant steps in OPT arise
due to the construction of maximal consistent sets of columns that
are not coverings (do not satisfy condition (a)).

The dualization algorithms RS and MMCS were proposed in
\cite{MU2011, MU2014}. Their description makes use of concepts of
hypergraph theory. These algorithms are based on constructing sets
of vertices of a hypergraph  $\cH$ satisfying the ``crit''
condition, which is equivalent to compatibility condition 2) for
the corresponding set of columns of the incidence matrix~of~$\cH$.
Thus, the approach proposed in \cite{MU2011, MU2014} for the
construction of dualization algorithms is not new (in fact, RS and
MMCS are asymptotically optimal algorithms).

The algorithm RUNC-M \cite{DP2015} is one of the fastest among
asymptotically optimal algorithms. As a rule, RUNC-M is less
time-consuming than the asymptotically optimal algorithms
constructed in \cite{D1977, D2004, DI2008, D1987, DZ1997, DZ2000,
    D2003, DP2014, MU2011, MU2014}. In this paper, a new
implementation of RUNC-M is developed. This implementation works
on a number of test tasks significantly faster than the
implementation described in \cite{DP2015}.

Due to the complexity of dualization, the use of parallel
computations is essential. In the development of parallel
dualization algorithms, the focus is on deriving theoretical worst
case complexity. However, such estimates can be obtained only for
some special cases of dualization. (see \cite{KBGE2007}).

In this paper, a new practical parallization scheme for
asymptotically optimal dualization algorithms is constructed. The
proposed scheme is of static nature and is based on statistical
estimations of subtasks size. There exist simple and obvious
practical parallelization schemes of asymptotically optimal
dualization algorithms. Their main disadvantage is an unbalanced
load of processors which produces insufficient speedup.

Let us describe the computational subtasks in question. Let $H$ be
an irreducible covering of the Boolean matrix $L$ consisting of
columns with indicies $j_1,\dots, j_r$, where $j_1<\dots<j_r$.
Then $H$ is called \emph{irreducible $j_1$-covering}. The $j$-th
computational subtask is to construct all irreducible
$j$-coverings of $L$. Therefore, we define the $j$-th subtask size
$\nu_j(L)$ as the ratio of the number of irreducible $j$-coverings
to the number of all irreducible coverings.
For optimal load balancing, one should know the values of
$\nu_j(L)$; however, they become known only after the dualization
is completed.

The proposed parallelization scheme is based on processing random
$r$-by-$n$ submatrices of the input matrix, where $r$ is a
parameter that doesn't exceed $m$.
The processor load is scheduled only after the calculation of the
subtask sizes for a given number of random submatrices.

The validity of estimating $\nu_j(L)$ based on random submatrices
is justified statistically.
First, we introduce a special random variable $\eta_r$ defined on
the set of $r$-by-$n$ submatrices and their irreducible coverings.
Its value is defined as the least index of columns in the covering.
Next, we test the statistcal hypothesis that the distribution of
$\eta_r$ is determined by the subtask sizes of the dualization of
the matrix $L$.
It is found that, according to the Chi-squared test, this
hypothesis can be accepted with confidence when $r \ge m/2$.

The scheme is highly scalable (a balanced load and almost maximal
speedup). In this paper, the proposed scheme is applied to the
algorithm RUNC-M. However, it is also applicable for all
dualization algorithms that sequentially construct sets of
irreducible $1$-coverings, $2$-coverings and so on.

The paper is organized as follows. In Section 2, we give a formal
definition of the asymptotically optimal dualization algorithm and
describe its basic structure via decision trees. In Section 3, we
describe the algorithm RUNC-M, provide some details about its new
implementation, and compare it experimentally with the previous
RUNC-M version from \cite{DP2015}. Our approach to parallelizing
asymptotically optimal algorithms is described in Section 4. This
approach is applied to the algorithm RUNC-M and tested in Section
5. Section 6 contains conclusions.

\section{Terms and definitions}

Let $M_{mn}$ be a set of $m{\times}n$ Boolean matrices and
$P_{mn}(X) = |X| / |M_{mn}|$ for $X \subseteq M_{mn}$. It is said
that $f(L)\approx g(L)$, $m, n \to \infty$, for almost all $L\in
M_{mn}$ if
$$
\forall \delta>0, \quad \exists \lim_{m,n \to \infty} P_{mn} \left(\left\{L \colon
\left |1-f(L)g(L)^{-1} \right|<\delta \right\} \right) = 1.
$$

Let us consider the following class of algorithms for enumerating
the irreducible coverings of a Boolean matrix $L\in M_{mn}$. Each
algorithm $A$ in this class constructs a finite sequence $Q_A(L)$
of column sets of $L$ that contains all elements from $\cP(L)$. It
is assumed that some elements of $Q_A(L)$ can be repeated. At each
step, the algorithm $A$ constructs an element of $Q_A(L)$ and
checks whether it belongs to $\cP(L)$. If the constructed element
is in $\cP(L)$, then $A$ additionally verifies in a polynomial
time whether it was earlier constructed. Let $N_A(L)$ be a number
of steps of the algorithm $A$ (length of $Q_A(L)$).

The algorithm  $A$ is asymptotically optimal with a polynomial time delay  $d$ if
\begin{itemize}
    \item
    $d$  is bounded above by a polynomial in $m$ and $n$;
    \item
    each step in  $A$ consists of at most  $d$ elementary operations (one matrix element
    access);
    \item
    $N_A(L)\approx |\cP(L)|$, $m,n \to \infty$, for almost all $L\in M_{mn}$.
\end{itemize}

Let  $\cS(L)$ be the set of all identity submatrices of the matrix
$L$. The number of maximal consistent sets of columns is bounded
above by $|\cS(L)|$. The theoretical substantiation of
asymptotically optimal dualization algorithms is based on the
following statement.
If $m \le n^{1-\varepsilon}$, where $ \varepsilon > 0$, then $|\cS(L)| \approx
|\cP(L)|$,
$m, n \to \infty$, for almost all $L\in M_{mn}$ (see \cite{D1977}).

The column  $j$ is said to cover the row  $i$ of a matrix  $L$ if
$a_{ij} = 1$. Let  $H$ be a set of columns of $L$ . The set  $H$
is said to cover the row $i$ if there exists a $j \in H$ covering
$i $. Let the set of columns $H$ be consistent. The column $j$ of
$L$ is said to be \emph{compatible} with the set $H$ if set
$H\cup\{j\}$ is consistent;
otherwise this column is called \emph{incompatible} with the set
$H$.

The work of an asymptotically optimal dualization algorithm  can
be regarded as a unidirectional traversal of the branches of a
decision tree. Each tree vertex is associated with the tuple
$(H,R,C)$, where $H$ is the set of columns of $L$, and $R$ and $C$
are, respectively, the sets of rows and columns describing the
submatrix of $L$, and $C$ and $H$ are disjoint. The vertex
$(\emptyset, R_0, C_0)$, where $R_0$ and $C_0$ describe the whole
matrix $L$, is the tree root. The leaf vertices are either
irreducible coverings or correspond to redundant steps of the
algorithm. Every step of the algorithm represents a transition
from one terminal vertex (or root) to another one. A transition
from one internal vertex to the next one is performed by adding a
column of $L$ to the set $H$. It is assumed that the number of
elementary operations at each step is polynomially bounded in $m$
and $n$.

Asymptotically optimal algorithms can be classified into two
types. Among the first type are the algorithms enumerating the
maximal identity submatrices of $L$. Such algorithms execute
redundant steps in which solutions constructed in the preceding
steps are constructed once more. Examples of such algorithms are
AO1 \cite{D1977} and AO2 \cite{D2004}. The algorithms of the
second type  are based on the enumeration of maximal consistent
sets of columns. This class includes the algorithms OPT
\cite{DI2008}, MMCS, RS \cite{MU2011, MU2014}, PUNC, and RUNC-M
\cite{DP2015}.

\section{Algorithm RUNC-M}

The algorithm RUNC-M is described as a recursive procedure
RUNCM. 
The first call RUNCM($L,H_0,R_0,C_0$) should be done with the parameters $H_0=\emptyset$, 
$R_0=\{1,\ldots,m\}$, $C_0=\{1,\ldots,n\}$. Notice that the parameters are passed by
value.

\begin{AlgProcedure}{RUNCM($L,H_0,R_0,C_0$)}
    \STATE $C_0^{min}=\{j\in C_0 | a_{ij}=1\}$, where $i\in R_0$ is the index \\of the row
    with the least sum $\sum_{j\in C_0} a_{ij}$;
    \FORALL{$j \in C_0^{min}$}
    \STATE $R   \leftarrow R_0$
    \STATE $C_0 \leftarrow C_0 \setminus \{j\}$
    \STATE $C   \leftarrow C_0$
    \STATE $H   \leftarrow H_0 \cup \{j\}$
    \STATE Eliminate from $R$ the rows that are covered by column $j$
    \IF{$R = \emptyset$}
    \STATE Save the set of columns $H\in \cP(L)$
    \ELSE
    \STATE Eliminate from $C$ the columns that are incompatible with $H$
    \STATE \CALL RUNCM($L,H,R,C$)
    \ENDIF
    \ENDFOR
\end{AlgProcedure}

The following criterion is used for incompatible columns
elimination. A row $i$ of the matrix $L$ is called
\emph{supporting} for $(H, j)$, $j \in H$, if $a_{ij}=1$ and
$a_{il} = 0$, $l \in H \setminus \{j\}$. The set of supporting
rows for $(H, j)$, $j \in H$, is denoted by $S(H,j)$. A column $u$
is compatible with $H$ if and only if  there is no column  $j \in
H$ such that column $u$ covers all rows from $S(H,j)$.

We developed a new implemetation of the algorithm RUNC-M, which is
available at
(\url{https://github.com/ankifor/dualization-OPT.git}).

\section{Parallelizing asymptotically optimal algorithms}

In this section we describe a practical parallelization S-scheme
that computes the relative subtask sizes by estimating the values
$\nu_j(L)=|P_j(L)| \big/ |P(L)|,\,j\in \{1, \ldots, n\}$.
The proposed scheme is is designed for processing the Boolean
matrices in which the number $m$ of rows is significantly greater
than the number $n$ of columns.

Let $L\in M_{mn}$ and $r\leq m$.
The set of all $r$-subsets of $\{1, \ldots, m\}$ is denoted by $W_m^r$.
Let $w \in W_m^r$; then $L^w$ denotes the submatrix of $L$
consisting of the rows of $L$ with indicies from $w$.
We define a function $\eta_r$ acting from $\Omega_r = \{ (L^w,H) : w\in W_m^r, H\in 
P(L^w)\}$ to $\{1, \ldots, n\}$ such that $\eta_r(L^w, H)$ equals $j$ if $H\in P_j(L^w)$.

We choose $t$ random submatrices $L^{w_1}, \ldots, L^{w_t}$, $w_s \in W_m^r$, $s \in \{1, 
\ldots, t\}$ and build $P(L^{w_s})$ for each of them.
Then we take $u$ random irreducible coverings $H_1^s, \ldots, H_u^s$ from these sets.
Next, we compose a sample $\vec{x}=(x_1,\ldots,x_N)$, $N = t\cdot u$, of values of  
$\eta_r(L^{w_s}, H_v^s), \; s\in\{1, \ldots, t\}, \; v \in \{1, \ldots, u\}$,
and calculate the frequency $f^*_r(j)$ of occurrence of $j$ in $\vec x$.
The quantity $f^*_r(j)$ is used as an estimation of $\nu_j(L),
j\in \{1, \ldots, n\}$.

A statistical justification of this approach is given below. We
also find out the values of $r$ under which the resulting
estimates are sufficiently accurate.
On the one hand, the integer $r$ should be as small as possible to
reduce the computation time of $f^*_r(j)$. On the other hand,
these estimations should be sufficiently reliable.

Let $\Omega_r$ be a sample space.
The probability of event $(L^w,H)$ is set to $\left(\binom{m}{r} |P(L^w)|\right)^{-1}$.
Then we denote the probability of event $\eta_r(L^w,H)=j$ by $f_r(j)$.

To test the statistical hypothesis $H_0: f_r(j)=\nu_j (L)$ about
the distribution of the random variable $\eta_r$, we use the
Chi-squared test with the statistic
\[
Z_r(\vec{x})= N \sum\limits_{j=1}^{n}{\frac{\left( f^*_r(j)-\nu_j(L)\right)^2}{\nu_j(L)}}.
\]
P-value is denoted by $\gamma_r^*(\vec{x})=1-\chi_{n-1}^2
(Z_r(\vec{x}))$, where $\chi_{n-1}^2$ denotes the cumulative
chi-squared distribution function with $(n-1)$ degrees of freedom.
Small values of $\gamma_r^*(\vec{x})$ argue for rejecting $H_0$.

Now we conduct an experiment.
Generate $20$ random $m$-by-$n$ matrices.
Then dualize these matrices and calculate the exact values of
$\nu_j(L)$, $j \in \{1, \ldots, n\}$.
Let $t=20$ and $u=50$. For each matrix $L$ and for each $r$, $r
\in \{10,13,15,18,20,25,30,35\}$ such that $r < m$, construct a
sample $\vec x$ from the values $\eta_r$ and calculate the
statistic $Z_r(\vec{x})$ and p-value $\gamma_r^*(\vec{x})$.

The median values of $Z_r(\vec{x})$ and $\gamma_r^*(\vec{x})$ are
presented in Table \ref{tab:z_r} for each configuration
$30{\times}150$, $40{\times}120$, $50{\times}100$, $70{\times}70$
and different values of $r$.
According to Table \ref{tab:z_r}, $\gamma_r^*(\vec{x})$ becomes
significant at $r \ge m/2$. That is why a further increase of $r$
will not make the approximation of $\nu_j(L)$ much more accurate.

\begin{figure}[t]
    \centering
    \includegraphics[width=0.8\linewidth]{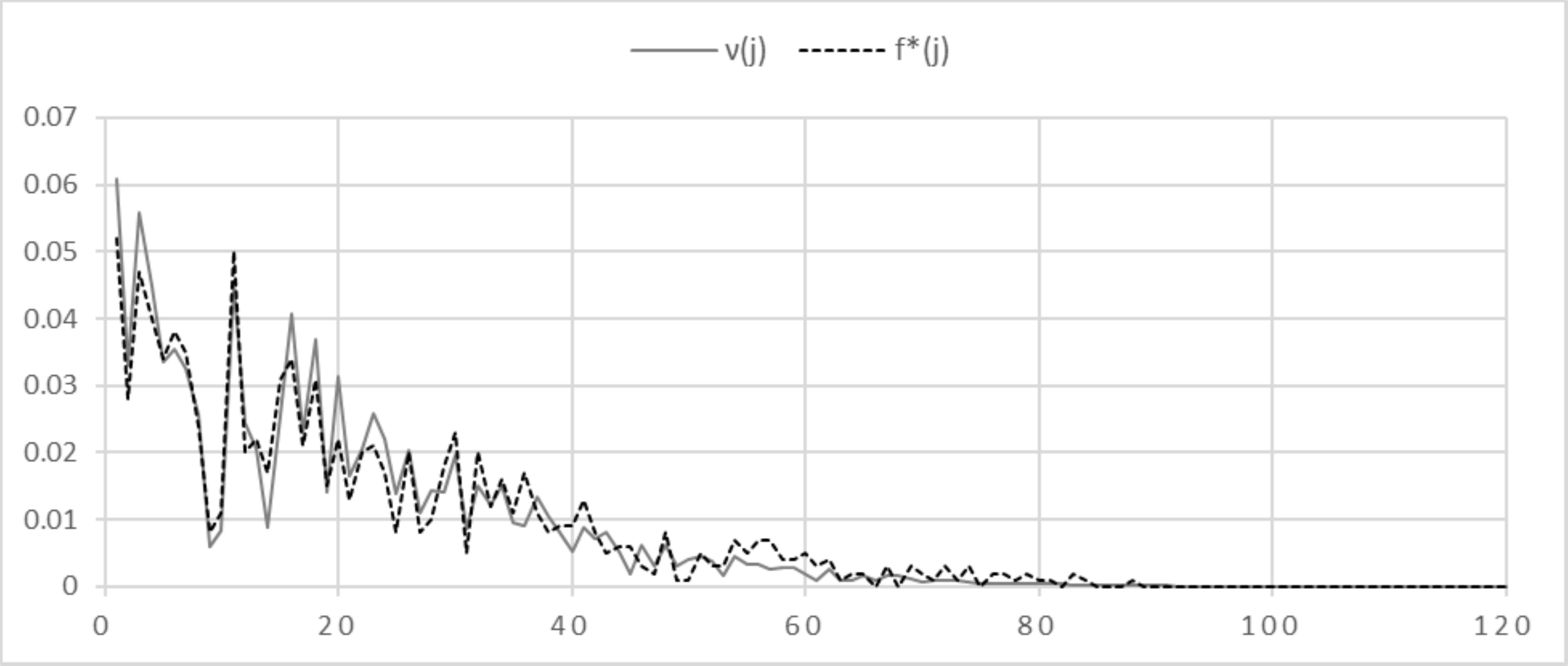}
    \caption{Plots of $\nu_j(L)$ and $f_r^*(r)$ as functions of $j$, where $m=30,
        n=120, and r=15$}
    \label{fig:nu_j}
\end{figure}

Consider the configuration $30{\times}150$ as an example. At
$r=15$, we observe the so-called ``phase transition'' --- the
function $Z_r(\vec{x})$ stabilizes after this point.
The plots of $\nu_j(L)$ and $f_r^*(j)$ for this case are presented
in Fig. \ref{fig:nu_j}.

To sum up, the experiment shows that $\nu_j(L)$ can be used as an
estimate of $f_r(j)$ at $r \ge m/2$.
Moreover, the frequency $f^*_r(j)$ is well-known to be a ``good''
estimate for the probability $f_r(j)$.
Thus, $f^*_r(j)$ can be used as an approximation of $\nu_j(L)$
under the conditions stated above.

\begin{table}[t]
    \caption{Median values of $(Z_r(\vec{x}), \gamma_r^*(\vec{x}))$ for chi-squared test.}
    \label{tab:z_r}
    \centering\medskip
    \begin{tabular}{l|l|l|l|l}
        \hline
        $r \setminus m{\times}n$
        & $30{\times}120$
        & $40{\times}120$
        & $50{\times}100$
        & $70{\times}70$ \\
        \hline
        $10$ & $(159, <10^{-4})$ & $(167, <10^{-4})$ & $(235, <10^{-4})$ & $(382, 
        <10^{-4})$ \\
        $13$ & $(99, <10^{-4}) $ & $(132, <10^{-4})$ & $(157, <10^{-4})$ & $(234, 
        <10^{-4})$ \\
        $15$ & $(77, 0.0134)$ & $(112, <10^{-4})$ & $(117, <10^{-4})$ & $(187, <10^{-4})$ 
        \\
        $18$ & $(74, 0.028) $ & $(90, 0.0002)$ & $(96, <10^{-4}) $ & $(147, <10^{-4})$ \\
        $20$ & $(60, 0.0815)$ & $(63, 0.0546)$ & $(89, <10^{-4}) $ & $(131, <10^{-4})$ \\
        $25$ & $(54, 0.315) $ & $(60, 0.0876)$ & $(50, 0.1382)$ & $(85, <10^{-4}) $ \\
        $30$ & $-           $ & $-           $ & $-           $ & $(68, 0.0001)$ \\
        $35$ & $-           $ & $-           $ & $-           $ & $(54, 0.0478)$ \\
        \hline
    \end{tabular}
\end{table}

After computing the estimates $f_r^*(j)$, we can proceed to scheduling the processor load.
Let we have $p\leq n$ processors and let the $j$-th subtask be executed by the processor 
with the index $N_j$.
The vector $\vec{N^p}=(N_1,\dots,N_n)$ is called \emph{schedule}.
The \emph{load balance} of the $k$-th processor is defined as
\[
\sigma_k(\vec{N^p}) = \sum\limits_{j\in J_n: N_j=k}{\nu_j(L)}.
\]
To obtain an efficient schedule, the following optimization problem should be solved.
\begin{equation}\label{eq:sigmamin}
\sigma(\vec{N^p}) = \max_{k \in J_p} {\sigma_k(\vec{N^p})} \to \min_{\vec{N^p}}.
\end{equation}

We propose the following procedure \Proc{DistributeTasks} for finding an approximate 
solution to problem \ref{eq:sigmamin},
which is based on a greedy strategy.
The parametres of the procedure are the number $p$ of processors, the number $n$ of 
columns of $L$, and the
vector $\vec f^*_r = (f^*_r(1), \dots, f^*_r(n))$ of estimators for $\nu_j(L)$.

\begin{AlgProcedure}{\Proc{DistributeTasks}($p, n, \vec f^*_r$) $\rightarrow
        (\vec{N^p},\sigma)$}
    \FORALL{$k \in \{1,\dots, p\}$}
    \STATE $\sigma_k \leftarrow 0$
    \ENDFOR
    \FOR{$j \in \{1,\dots, n\}$}
    \STATE $k_0 \leftarrow \argmin_{1 \le k \le p}{\sigma_k}$
    \STATE $N_j \leftarrow k_0$
    \STATE $\sigma_k \leftarrow \sigma_k + f^*_r(j)$
    \ENDFOR
\end{AlgProcedure}

\section{Parallel RUNC-M Test Results}

Testing was performed on the supercomputer IBM Blue Gene/P of the
Lomonosov Moscow State University.

Each computation node contains four PowerPC 450 processor cores
running at 850 MHz, 2 GB DRAM, and communication interfaces.
Computations were performed in the virtual-node mode (four MPI
processes per node, 1GB limit per process; cannot create
additional threads).

Let $p$ be the number of processors and $T_k(p)$ be the algorithm
execution time (in seconds) on the $k$-th processor. Let
$T(p)=\max_k T_k(p)$ and $T_\Sigma(p)=\sum_k T_k(p)$. The number
$s_k(p)=(T_k(p)) \big/ (T_\Sigma(p) )$ is called the realized load
level of the $k$-th processor. The following measures are of
interest:
\begin{enumerate}
    \item
    Algorithm speedup $S(p)=T(1)/T(p)$ ;
    \item
    Load balance uniformity $E(p)=S(p)/p$;
\end{enumerate}

The speedup $S(p)=p$ at $p\geq 1$ is almost maximal. If $E(p)$ is
close to unity, then the load balance is considered to be uniform.
The measure $s(p)$ is an analog of $\sigma(\vec{N^p})$ (see formula
\ref{eq:sigmamin}).

\begin{figure}[t]
      \begin{subfigure}{0.49\textwidth}
    \centering
    \includegraphics[width=0.8\linewidth]{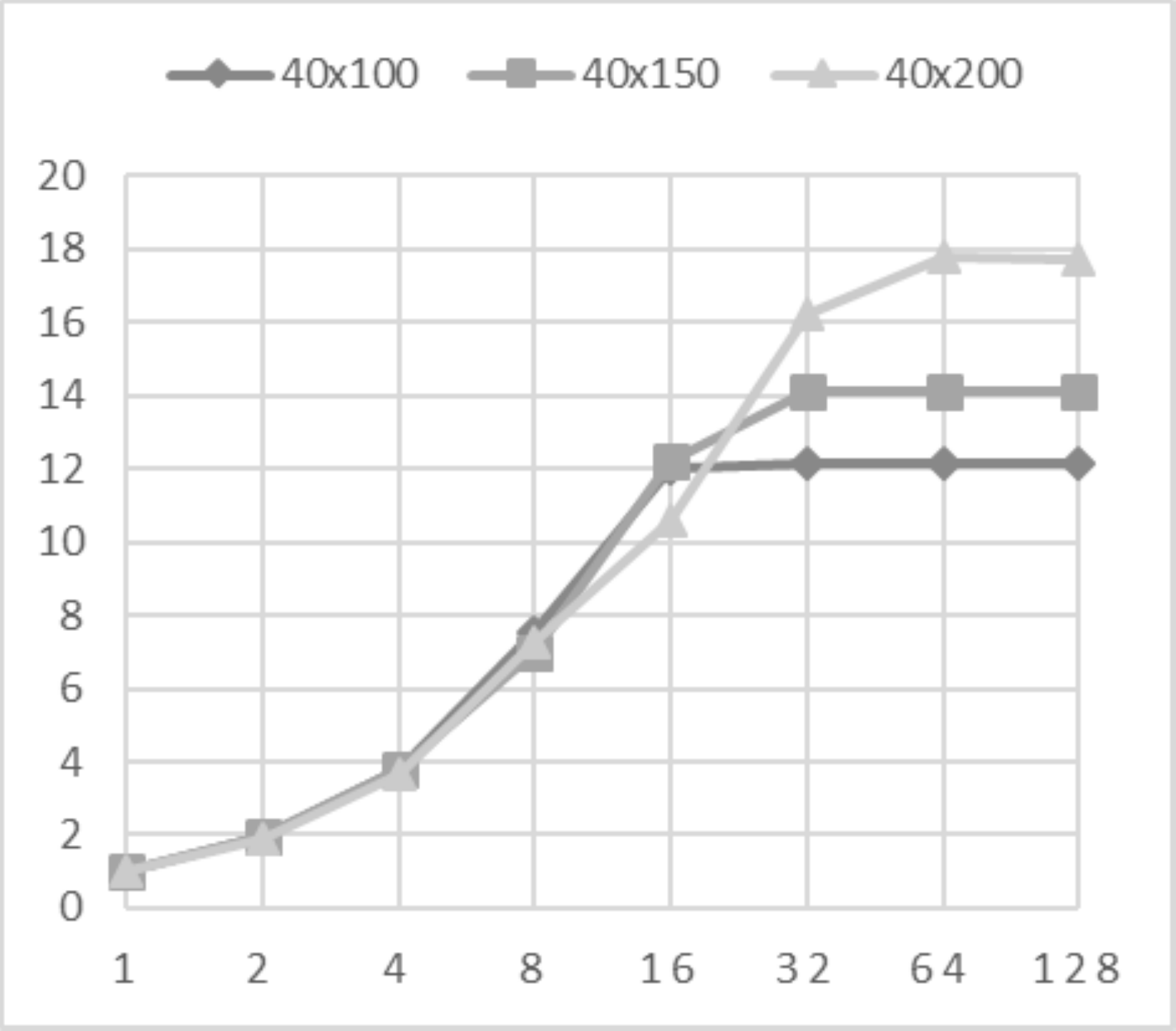}
    \caption{Plot of $S(p)$ as function of $p$}
       \end{subfigure}
       \begin{subfigure}{0.49\textwidth}
    \centering
    \includegraphics[width=0.8\linewidth]{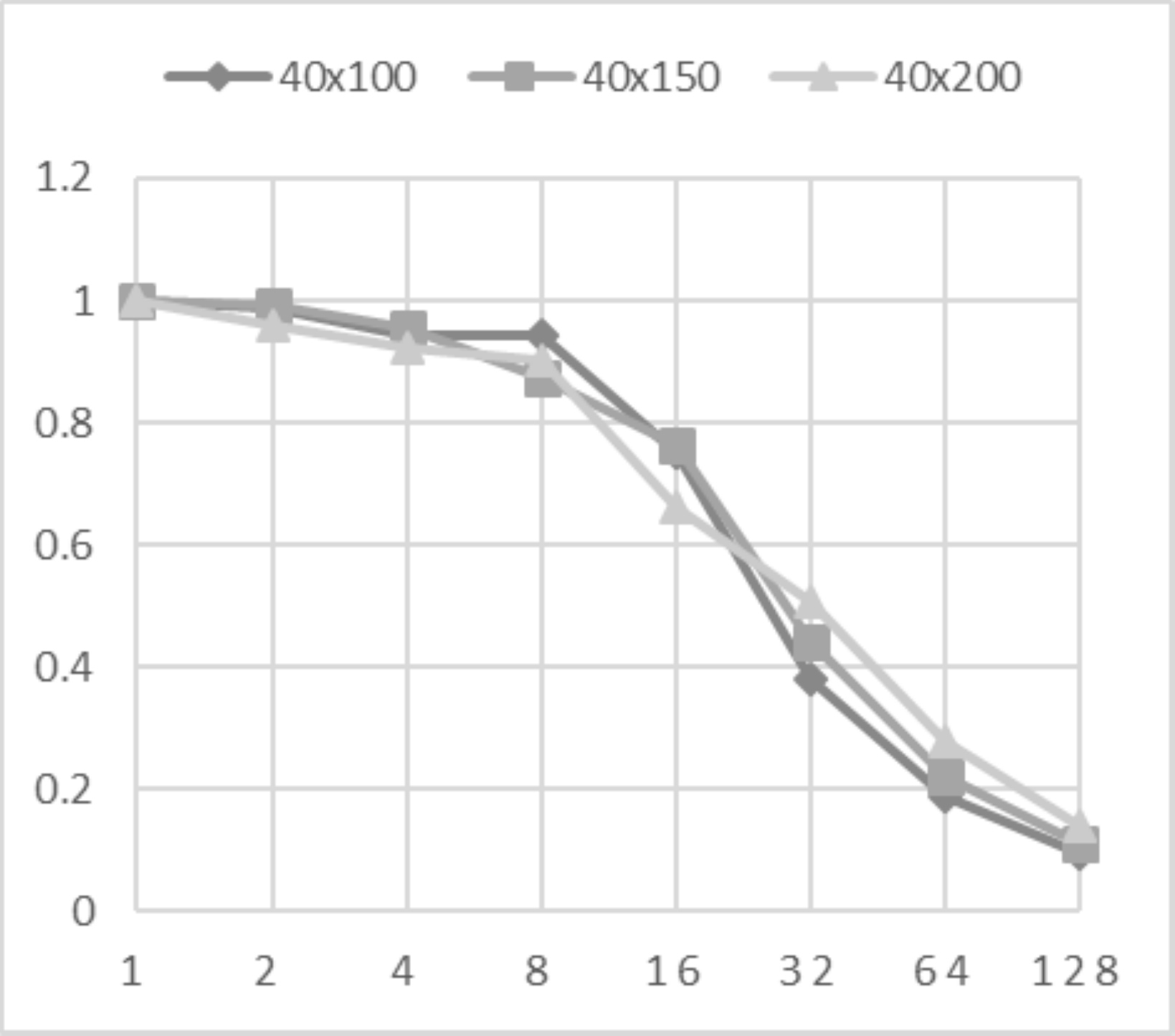}
    \caption{Plot of $E(p)$ as function of $p$}
       \end{subfigure}
    \centering
    \caption{
        Strong scalability of the parallel RUNC-M version
        when $m=40$ and $n\in\{100,150,200\}$}
    \label{fig:s_scheme1}
\end{figure}

\begin{figure}
       \begin{subfigure}{0.49\textwidth}
    \centering
    \includegraphics[width=0.8\linewidth]{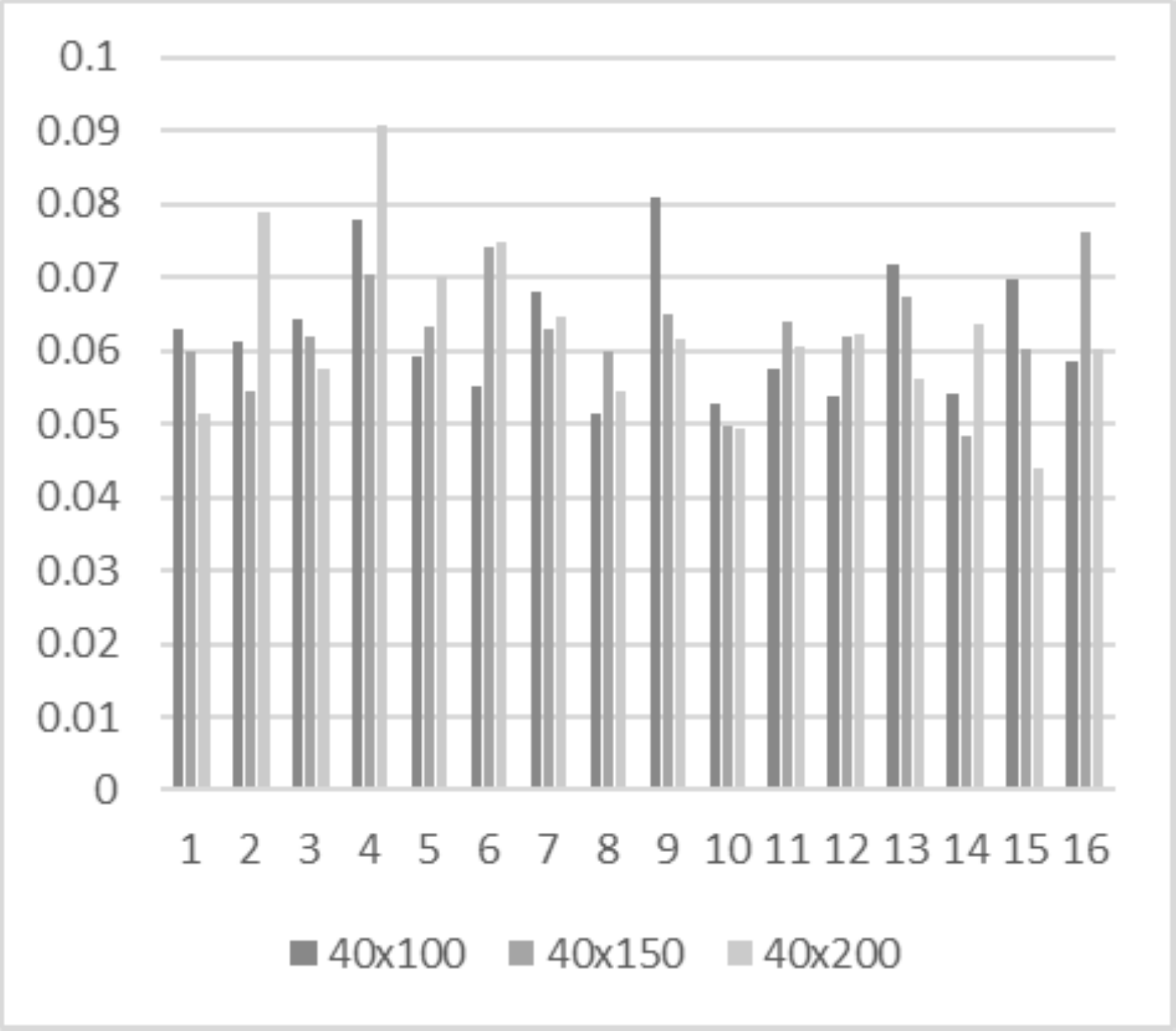}
    \caption{Plot of $s_k(16)$ as function of $k$}
       \end{subfigure}
       \begin{subfigure}{0.49\textwidth}
    \centering     
    \includegraphics[width=0.8\linewidth]{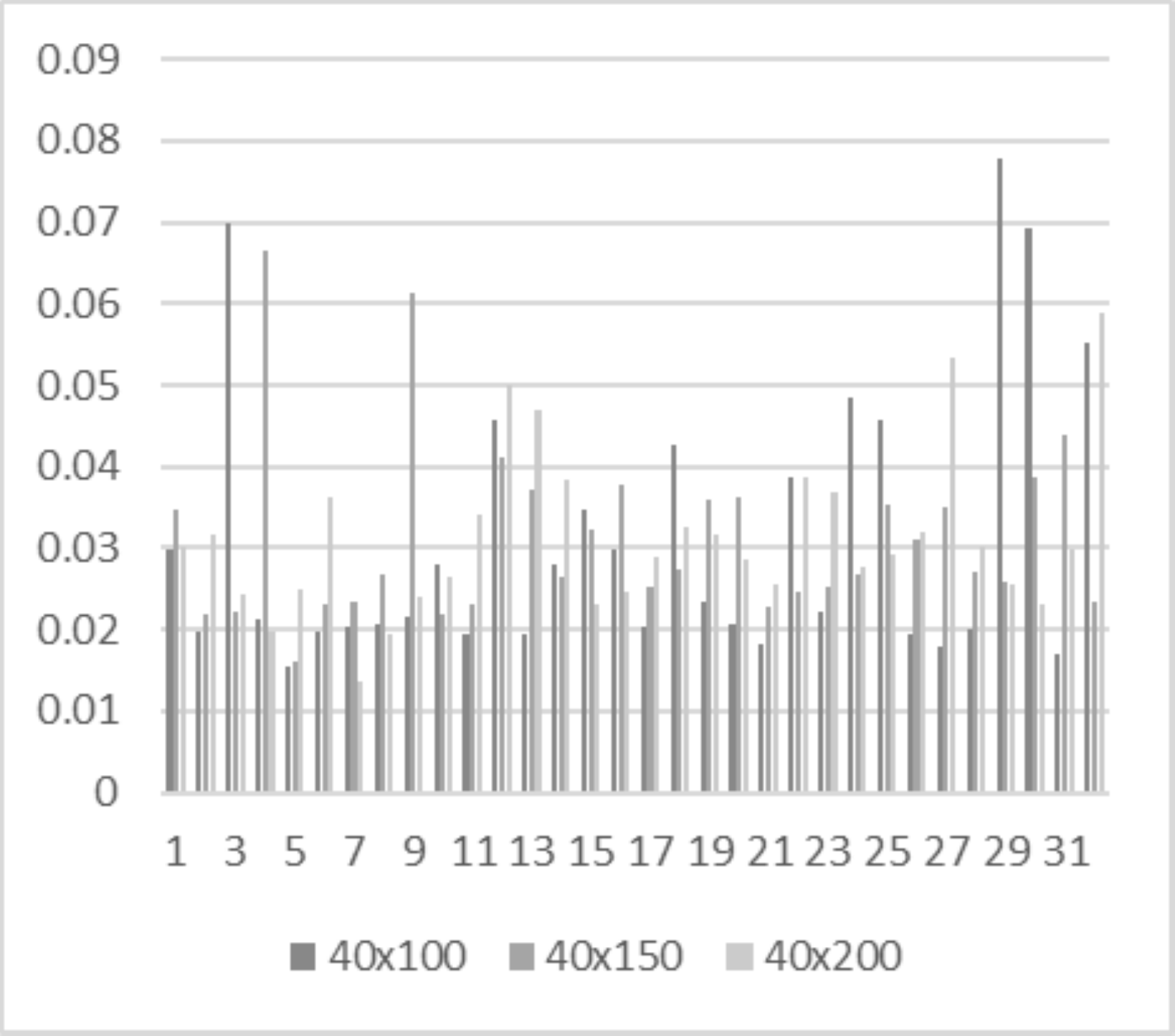}
    \caption{Plot of $s_k(32)$ as function of $k$}
       \end{subfigure}
    \centering
    \caption{
        Realized load level of the parallel RUNC-M version
        when $m=40$ and $n\in\{100,150,200\}$}
    \label{fig:s_scheme2}
\end{figure}

\begin{table}[h]
    \scriptsize
    \caption{Parallel RUNC-M execution time (sec) depending on the number of processors}
    \label{tab:results2}
    \centering
    \medskip
    \begin{tabular}{c|cccccccc}
        Data / $p$ & 1 & 2 & 4 & 8 & 16 & 32 & 64 & 128\\
        \hline
        \hline
        $U(30,100)$ & $3.95$ & $2.03$ & $1.05$ & $0.59$ & $0.37$ & $0.32$ & $0.32$ & 
        $0.32$ \\
        $U(30,150)$ & $39.1$ & $20.0$ & $10.4$ & $5.21$ & $3.46$ & $2.32$ & $2.33$ & 
        $2.32$ \\
        $U(30,200)$ & $231 $ & $116 $ & $61.5$ & $32.2$ & $18.8$ & $13.8$ & $13.8$ & 
        $13.8$ \\
        $U(40,100)$ & $11.5$ & $5.83$ & $3.05$ & $1.53$ & $0.96$ & $0.95$ & $0.95$ & 
        $0.95$ \\
        $U(40,150)$ & $133 $ & $67.1$ & $34.8$ & $19.1$ & $10.9$ & $9.44$ & $9.43$ & 
        $9.43$ \\
        $U(40,200)$ & $654 $ & $328 $ & $177 $ & $90.5$ & $61.8$ & $40.4$ & $36.8$ & 
        $36.8$ \\
        \hline
    \end{tabular}
\end{table}

Experiments were carried out on random $m$-by-$n$ matrices,
where $m\in \{30,40\}$ and $n\in \{100,150,200\}$. The parameter
$r$ of the statistical procedure discribed above is equal to
$m/2$. In order to estimate the values of $\nu_j(L)$, the
dualization problem is solved for $r$-by-$n$ submatrices $L_w$ of
the matrix $L$. The computation results are presented in Table
\ref{tab:results2}; they include the execution time of the
parallel algorithm on different number of processors. The plots of
$S(p)$ and $E(p)$ are given in Fig. \ref{fig:s_scheme1} for the
case when $m=40$ and $p\in\{1,2,4,\ldots,128\}$.

As can be seen in these figures and table, the parallel RUNC-M
version has almost linear speedup $S(p)$ and loads processors in a
balanced way when the number of processors $p$ is below some
threshold $p^*$. Generally speaking, this threshold depends on the
dualized matrix size. For example, $p^*$ equals $32$ when $m=40,
n=200$, and it equals $16$ when $m=40, n=100$. When the number of
processors is greater than the threshold, the execution time
$T(p)$ stops to improve. This is because parallelization takes
place on the first level of the decision tree built by the
algorithm RUNC-M. Under the proposed approach, the size of the
computational subtasks are significantly different. Therefore, it
 is here immposible to distribute tasks in a balanced manner over a large
number of processors.

The realized load levels $s_k(p)$ for each processor are presented
in Fig. \ref{fig:s_scheme2}. As seen in this figure, some realized
load levels differ by several fold when $m = 40, n = 200$ and
$p=32$. That might be due to insufficient quality of the subtask
size estimates $f_r^*(j)$ or non-optimality of the task
distribution schedule. Nevertheless, the variance of the realized
load levels is fairly small for $p=16$, which is in agreement with
the high speedup.

\section{Conclusion}

In this paper an approach \cite{DNP2014} to the parallel algorithm
construction for discrete enumeration problems is developed.
This approach is based on statistical estimates of computational
subtask sizes.
Subtasks are assigned to processors in accordance with
precalculated schedule.
To construct this schedule, the distribution of a special random
variable used for estimating the subtask sizes is found. Then, the
load balance of processors is optimized.
A novel efficient parallelization scheme for asymptotically
optimal dualization algorithms based on the proposed approach is
developed.
The scheme is applied to the algorithm RUNC-M \cite{DP2015}, which
is the fastest known dualization algorithm.
The proposed approach to the the construction of parallel
dualization algorithms ensures high accuracy of subtask size
estimates, which under certain conditions leads to highly
efficient parallel algorithms.
However, the proposed approach is not that efficient when the
number of processors is large because the sizes of computational
subtasks can vary significantly (the parallelization is performed
at the first level of the decision tree built by the
asymptotically optimal dualization algorithm).

\section*{References}

\def\BibAuthor#1{\emph{#1}}
\def\BibTitle#1{{#1}}
\def\BibUrl#1{{\small\url{#1}}}


\end{document}